\documentclass[12pt,preprint]{aastex}
\shorttitle{Constraining the Emission Properties of TeV Blazar H1426+428}
\shortauthors{Kato, Kusunose, \& Takahara}

\begin{document}

\title{Constraining the Emission Properties of TeV Blazar H1426+428
by the Synchrotron-Self-Compton Model}

\author{Tomohiro Kato}
\author{Masaaki Kusunose}
\email{kusunose@kwansei.ac.jp}
\affil{Department of Physics, School of Science and Technology,
Kwansei Gakuin University, Sanda 669-1337, Japan}

\and

\author{Fumio Takahara}
\affil{Department of Earth and Space Science,
Graduate School of Science, Osaka University,
Toyonaka 560-0043, Japan}
\email{takahara@vega.ess.esi.osaka-u.ac.jp}


\begin{abstract}
H1426+428 is one of blazars that are observed by $\gamma$-rays in 
the TeV region.  Because TeV $\gamma$-rays from distant sources
are subject to attenuation by the extragalactic background light (EBL)
via electron-positron pair production, the intrinsic spectrum of the
TeV $\gamma$-rays should be inferred by using the models of radiation processes
and EBL spectrum.  We set constraints on the physical condition of 
H1426+428 with the synchrotron-self-Compton model applying several EBL models.
We find that the emission region of H1426+428 is moving toward us
with the bulk Lorentz factor of  $\sim 20$ 
and that its magnetic field strength is $\sim 0.1$ G.
These properties are similar to other TeV blazars 
such as Mrk 421 and Mrk 501.
However, the ratio of the energy density of nonthermal electrons to
that of the magnetic fields is about 190 and fairly larger than those of
Mrk 421 and Mrk 501, which are about 5 -- 20.
It is also found that the intensity of EBL in the 
middle and near infrared wavelengths should be low,
i.e., the intensity at 10 $\mu$m is about 
1 nW m$^{-2}$ sr$^{-1}$ to account for the observed TeV $\gamma$-ray flux.
Because the spectral data of H1426+428 in X-rays and $\gamma$-rays
used in our analysis were not obtained simultaneously,
further observations of TeV blazars are necessary to make 
the constraints on EBL more stringent.
\end{abstract}


\section{Introduction}

Recent observations of very high energy $\gamma$-rays revealed that
some blazars emit $\gamma$-rays in the TeV region
\citep[e.g.,][and references therein]{aha05b}.
The spectral energy distribution (SED) of these blazars has two peaks;
one in the X-ray region and the other in the very high energy $\gamma$-ray
region (GeV -- TeV).
The emission mechanisms of TeV blazars are thought to be synchrotron radiation
and inverse Compton scattering by nonthermal particles in 
relativistically moving jets \citep[e.g.,][]{mgc92,mt96,it96,lk00,kino02}.
Leptonic models that assume electrons and
positrons as emitting particles are most widely accepted.
Although hadronic models are viable,
which postulate emission from mesons and leptons produced
by the cascade initiated by proton-photon or proton-proton collisions
\citep[e.g.,][]{mb92,man93,muc03}, 
we do not consider these models in this paper.

It has been known that TeV $\gamma$-rays emitted by blazars
are attenuated by absorption due to the extragalactic background light (EBL) 
en route to the Earth, because $e^+ e^-$ pair production occurs 
via $\gamma + \gamma \rightarrow e^+ + e^-$.
The importance of the absorption of high energy $\gamma$-rays due to
cosmic background radiation was noticed quite early \citep{nik62,gs66,jel66}.
The effects of the attenuation on the SED of blazars were recognized by
\citet{sjs92} following EGRET observation of 3C 279 \citep{har92}.
Since then this effect has been extensively
studied by many authors \citep[e.g.,][]{ds94,ss96,kon03,kne04,dk05}.
The spectrum of EBL ($\lambda \sim 0.1$ to 1000 $\mu$m)
has been observed by various instruments \citep[][for review]{hd01,kas05}.
It is known that there are two peaks in the SED of EBL;
one at $\lambda \sim 1$ $\mu$m is associated with star light and the other
at $\lambda \sim 100$ $\mu$m is related with dust emission.
The wavelength region of $\sim 5$ to 60 $\mu$m is difficult to observe
because of interplanetary dust emission,
while the attenuation of TeV $\gamma$-rays is most effective
in this wavelength region
of EBL.  When a $\gamma$-ray with energy $E_\gamma$ collides 
with an EBL photon with energy $\varepsilon_{\mathrm{EBL}}$, 
pair production occurs
if $E_\gamma \varepsilon_{\mathrm{EBL}} \ge (m_e c^2)^2$ is satisfied, 
where $m_e$ and $c$ are the rest mass
of an electron and the light speed, respectively.
If $E_\gamma = 1$ TeV, a photon with 
$\lambda_\mathrm{EBL} = hc/\varepsilon_\mathrm{EBL} \lesssim 1$ $\mu$m
produces a $e^+ e^-$ pair, where $h$ is the Planck constant.

Because both the intrinsic SED of TeV blazars 
and the EBL spectrum are not well known, 
various attempts have been made to construct self-consistent models
of TeV blazars and the EBL spectrum. 
(Differences in various EBL models are discussed in detail by \citet{hd01}.)
For example, \citet{djs02} derived the intrinsic
$\gamma$-ray spectrum of Mrk 501 during 1997 high state, using the models
of EBL by \citet{ms01} (backward evolution model).
Recently \citet{dk05} employed a different way of constructing 
the models of the EBL spectrum.  
They adopted various observed EBL data and fitted the data with polynomials; 
12 different EBL template spectra with different
intensities in UV -- near IR, mid IR, and far IR spectra.
Using those spectra, they set constraints on the intrinsic spectra of
Mrk 421, Mrk 501, and H1426+428 and the acceptable range of the EBL spectrum
was estimated.

Most models of the SED of TeV blazars and the EBL spectrum 
have not taken into account the electron spectrum
in the emission region self-consistently.
Many authors assumed a given form of the energy spectrum of 
electrons or do not consider the emission mechanisms at all.
\citet{kon03}, on the other hand, 
solved kinetic equations of electrons and photons
simultaneously to calculate the emission spectrum of Mrk 501 and Mrk 421.
They calculated the optical depth for TeV $\gamma$-rays according to
the EBL models by \citet{ms01} and \citet{djs02}.
They obtained a fairly good fit to the observed spectrum of the blazars
with a rather large value of the Doppler factor $\cal D$ of the jet, 
i.e., ${\cal D} \sim 50$.

In this paper we make models of H1426+428 (at redshift $z = 0.129)$ 
and estimate the intensity of EBL, 
solving the kinetic equations of electrons
and photons to obtain the intrinsic SED of blazars,
so that the observed SED is consistent with the electron injection
and cooling processes.
X-ray observation of H1426+428 was carried out by \citet{cos01}
with {\it Beppo}SAX and X-ray spectrum was found to extend up to 100 keV.
\citet{fal04} observed H1426+428 with {\it RXTE} and found that
the peak energy in SED was sometimes in excess of $\sim 100$ keV
and at other times in the 2.4 -- 24 keV region.
TeV $\gamma$-rays have been observed by several groups
\citep{hor02,aha02,dja02,pet02,aha03}.
Because the redshift of H1426+428 is 0.129, 
the effects of the attenuation of TeV $\gamma$-rays
should be very large and the object can be a good test 
against various EBL models.

\citet{cos03} inferred the intrinsic spectrum of H1426+428 using
the model spectrum of EBL given by \citet{pri01},
who employed the semi-analytical models of galaxy formation,
as well as a EBL model spectrum designed to match the high fluxes
below 2 $\mu$m \citep{aha02}.
Their model spectrum of H1426+428 has a large value of
the ratio of Compton luminosity to synchrotron luminosity, 
$L_\mathrm{C}/L_s > 10$.
According to \citet{cos03}, external seed radiation fields for
inverse Compton scattering are necessary to account for 
$L_\mathrm{C}/L_s > 10$ and to avoid the Klein-Nishina effects in inverse
Compton scattering.
\citet{aha03} obtained a rising intrinsic spectrum in the TeV regime,
using the same EBL spectra as \citet{cos03} and \citet{ms01} adopted.

Since the estimate of the intrinsic SED of TeV blazars
largely depends on the EBL spectrum
and the uncertainty of the SED of EBL is quite large,
it is worth testing various possibilities of the EBL spectrum.
In this paper we use the templates of EBL derived by \citet{dk05},
because their model spectra cover a wider range in intensity
than those used by \citet{kon03}.
We present the method of calculations of the energy spectra of
photons and electrons in the jet and the spectrum of EBL 
in \S \ref{sec:model}.  Numerical results for energy spectra
of photons and electrons in the emission region of H1426+428 
are shown in \S \ref{sec:nr}.  
Finally summary of our results and discussion are given in \S \ref{sec:sum}.

\section{Model} \label{sec:model}

\subsection{Kinetic equations} \label{ssec:kinetic}

The emission mechanisms of H1426+428 are assumed to be
synchrotron radiation and synchrotron-self-Compton (SSC) scattering.
We calculate the SED of H1426+428 before the attenuation of $\gamma$-rays 
by simultaneously solving the kinetic equations of photons and electrons
in the emission region.  We use the same numerical scheme as used in
\citet{kus03} except that external soft photons are not included in this work.
We assume that the emission region is a spherical blob with
radius $R$ moving with a relativistic speed with Lorentz factor $\Gamma$.
It is also assumed that ${\cal D} = \Gamma$.
Nonthermal electrons are injected into the emission region 
at injection rate $q_e(\gamma)$ per unit $\gamma$ and unit volume
in the comoving frame of the blob, where $\gamma$ is the electron
Lorentz factor.  
The kinetic equation of nonthermal electrons is given by
\begin{equation}
\label{eq:kin-el}
\frac{\partial n_e(\gamma)}{\partial t}
+ \frac{n_e(\gamma)}{t_{e, \mathrm{esc}}}
= \frac{\partial}{\partial \gamma} \,
[(\dot{\gamma}_\mathrm{syn} + \dot{\gamma}_\mathrm{SSC})
n_e(\gamma)] + q_e(\gamma) \, ,
\end{equation}
where $t_{e, \mathrm{esc}}$ is the electron escape time from the emission
region,
and $\dot{\gamma}_\mathrm{syn}$ and $\dot{\gamma}_\mathrm{SSC}$
are the cooling rates by synchrotron radiation and SSC scattering, 
respectively.  Here $\dot{\gamma}_\mathrm{SSC}$ 
depends on the photon spectrum in the blob.
The injection rate of nonthermal electrons is given by
\begin{equation}
q_e(\gamma) = q_0 \gamma^{-p} \exp(-\gamma/\gamma_\mathrm{max}) \, ,
\quad \gamma \ge \gamma_\mathrm{min} \, ,
\end{equation}
where $p$, $\gamma_\mathrm{min}$, and $\gamma_\mathrm{max}$ are 
parameters, which are determined by fitting the emission spectrum with
observed one.

The kinetic equation of photon number spectrum
$n_\gamma(\varepsilon)$, where $\varepsilon$ is the photon energy, is given by
\begin{equation}
\label{eq:kin-ph}
\frac{\partial n_{\gamma}(\varepsilon)}{\partial t} 
+ \frac{n_{\gamma}(\varepsilon)}{t_{\gamma, \mathrm{esc}}} 
= \dot{n}_\mathrm{C}(\varepsilon)
+ \dot{n}_\mathrm{syn}(\varepsilon) - \dot{n}_\mathrm{abs}(\varepsilon) \, ,
\end{equation}
where $t_{\gamma, \mathrm{esc}} = R/c$ is the photon escape time from the 
emission region.
The photon production rate per unit volume in the blob 
by synchrotron radiation is denoted 
by $\dot{n}_\mathrm{syn}(\varepsilon)$.
The self-absorption by synchrotron radiation and the absorption
by $e^+ e^-$ pair production inside the blob are included in
$\dot{n}_\mathrm{abs}(\varepsilon)$.
The production and loss of photons by Compton scattering is given by
\begin{equation}
\dot{n}_{\rm C}(\varepsilon)
= - n_{\gamma}(\varepsilon) \, \int d\gamma \, n_e(\gamma) \,
R_\mathrm{C}(\varepsilon, \gamma) + 
\int\int d\varepsilon' \, d\gamma \, 
P(\varepsilon; \varepsilon', \gamma) \,
R_{\rm C}(\varepsilon', \gamma) \,
n_{\gamma}(\varepsilon') \, n_e(\gamma) \, ,
\end{equation}
where $R_\mathrm{C}(\varepsilon, \gamma)$ 
is the angle averaged scattering rate 
and $P(\varepsilon; \varepsilon', \gamma)$ is 
the probability that photon energy $\varepsilon'$
is changed to $\varepsilon$ by a single scattering with
an electron with $\gamma$ \citep{cb90}.
Here the exact Klein-Nishina cross section is used to calculate
$R_\mathrm{C}$ and $P$.
We solve equations (\ref{eq:kin-el}) and (\ref{eq:kin-ph}) simultaneously
to obtain the spectra of electrons and photons self-consistently.
The emission spectrum is calculated from the term of photon escape
in equation (\ref{eq:kin-ph})
and the observed spectrum is affected by the Doppler and beaming effects 
because of the relativistic motion of the emission region.
We include these effects to compare the model SED with observations.

%%%%%%%%%%%%%%%%%%%%%%%%%%%%%%%%%%%%%%%%%%%%%%%%%%%%%%%%%%%%%%%

\subsection{Optical Depth by EBL} \label{ssec:optdepth}

The interaction between high energy $\gamma$-rays
and EBL via $e^+ e^-$ pair production attenuates TeV $\gamma$-rays 
from blazars.  
The optical depth of the attenuation $\tau_{\gamma \gamma}$ is a function
of the $\gamma$-ray energy and the EBL spectrum.
For the calculation of $\tau_{\gamma \gamma}$ we assume a flat
$\Lambda$CDM cosmology with $H_0 =71$ km s$^{-1}$ Mpc$^{-1}$,
$\Omega_m = 0.27$, and $\Omega_\Lambda = 0.73$  \citep{ben03},
where $H_0$ is the Hubble constant and $\Omega_m$ and $\Omega_\Lambda$ are
the density parameters of matter and dark energy, respectively.
The EBL number density at $z$ is calculated as
\begin{equation}
\label{eq:ebl-local}
n_\mathrm{EBL}(\nu, z) = (1+z)^2 n_0(\nu_0)  \, ,
\end{equation}
where $n_0(\nu_0)$ is the EBL number spectrum measured at $z = 0$
and $\nu_0 = \nu (1+z)^{-1}$.  
Equation (\ref{eq:ebl-local}) assumes that the phase space density of
EBL is conserved.
The spectrum of EBL has been measured by various groups.
However, there is a large uncertainty in the observed spectra.
To take into account various possibilities, 
we follow the recently adopted method by \citet{dk05}.
They generated 12 EBL templates based on numerous observational data.
The templates are labeled as XYZ such as LLL, MLH, and LHH, etc.
Namely, X = L, M, or H representing the energy flux 
of the stellar component of EBL (L = low, M = medium, and H = high).
Also, Y = L or H, and Z = L or H, where Y denotes the flux level 
at 15 $\mu$m band and Z does at far-IR band.

The observed SED is calculated by multiplying
the model SED by $\exp(-\tau_{\gamma \gamma})$.
Because of this exponential factor of the attenuation,
the differences in $\tau_{\gamma \gamma}$ are significant in
construction of the SED models of blazars.

\section{Numerical Results}
\label{sec:nr}

\begin{figure}
\includegraphics[angle=0,scale=.80]{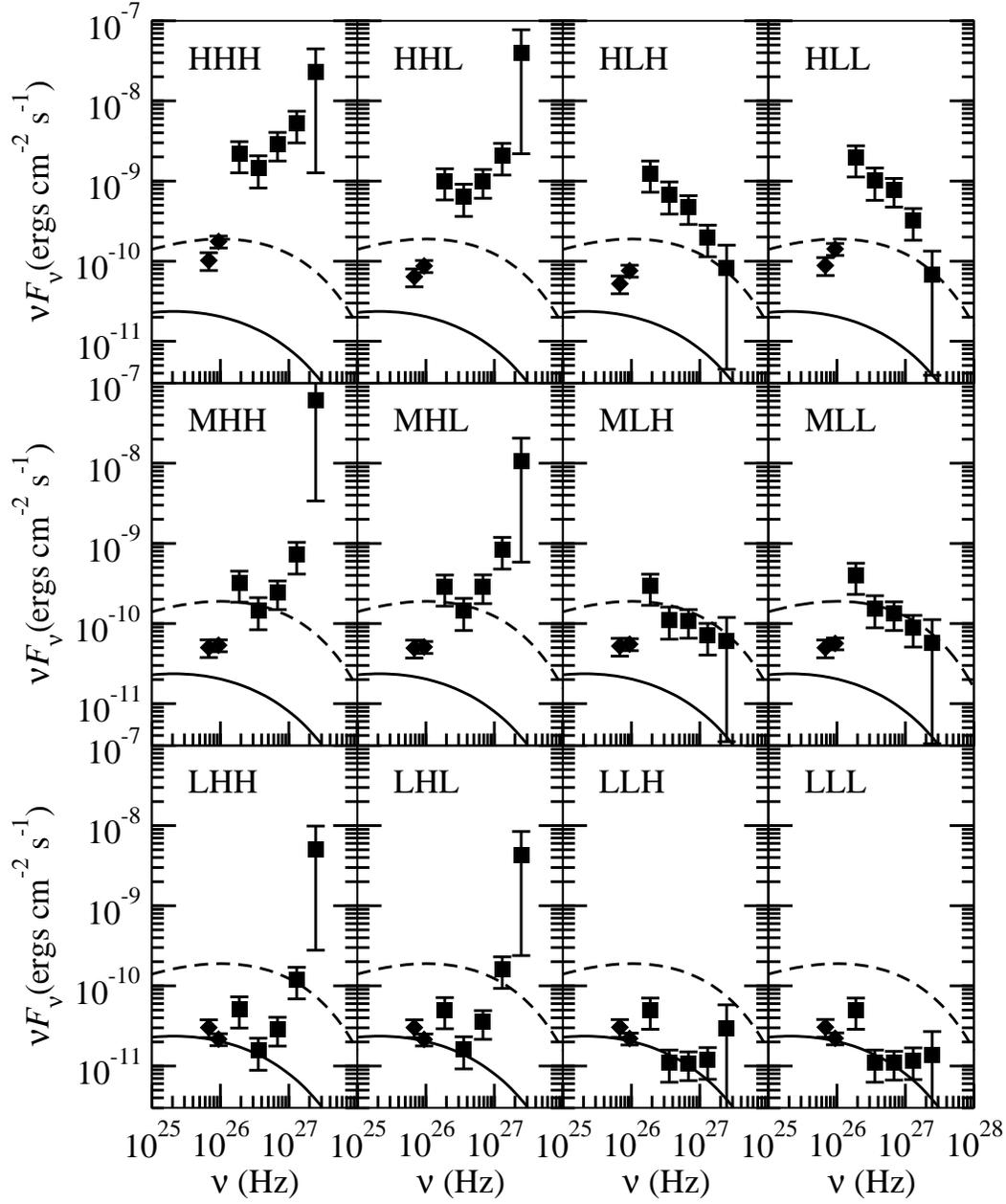}
\caption{TeV spectrum [data from Whipple \citep{hor02}]
and HEGRA [data from \citep{aha03}],
corrected by 12 different EBL models.
The solid lines are our model SED before the attenuation (base model),
which is shown by a dashed line in Fig. \ref{fig:h1426-sed}.
The dashed lines are also our model SED before the attenuation
shown by a dashed line in Fig. \ref{fig:h1426-XM-sed}.
\label{fig:h1426-all-EBL}}
\end{figure}

In Figure \ref{fig:h1426-all-EBL}, 
the $\gamma$-ray data obtained by the Whipple collaboration \citep{hor02}
and the HEGRA collaboration in the term 1999 -- 2000 \citep{aha03}
are shown.  These are calculated by multiplying the observed data
by $\exp (\tau_{\gamma \gamma})$ assuming various models of EBL.
It is found that the attenuation free SED does not depend much on Z of EBL,
i.e., SED is almost the same for Z = L and Z = H.
To distinguish the dependence on Z, $\gamma$-ray sources with higher
energies are needed.
When EBL models with Y = L are applied,
SED is flat or becomes downward as the photon
energy increases.
For EBL models with Y = H,
the peak of $\nu F_\nu$ before the attenuation may appear 
at photon energies greater than 50 TeV, 
which is hard to explain by SSC models; this will be discussed in
\S \ref{sec:sum}.
Because we search for the values of parameters with SSC models,
EBL models with Y = H are excluded.

\begin{figure}
\includegraphics[angle=0,scale=.80]{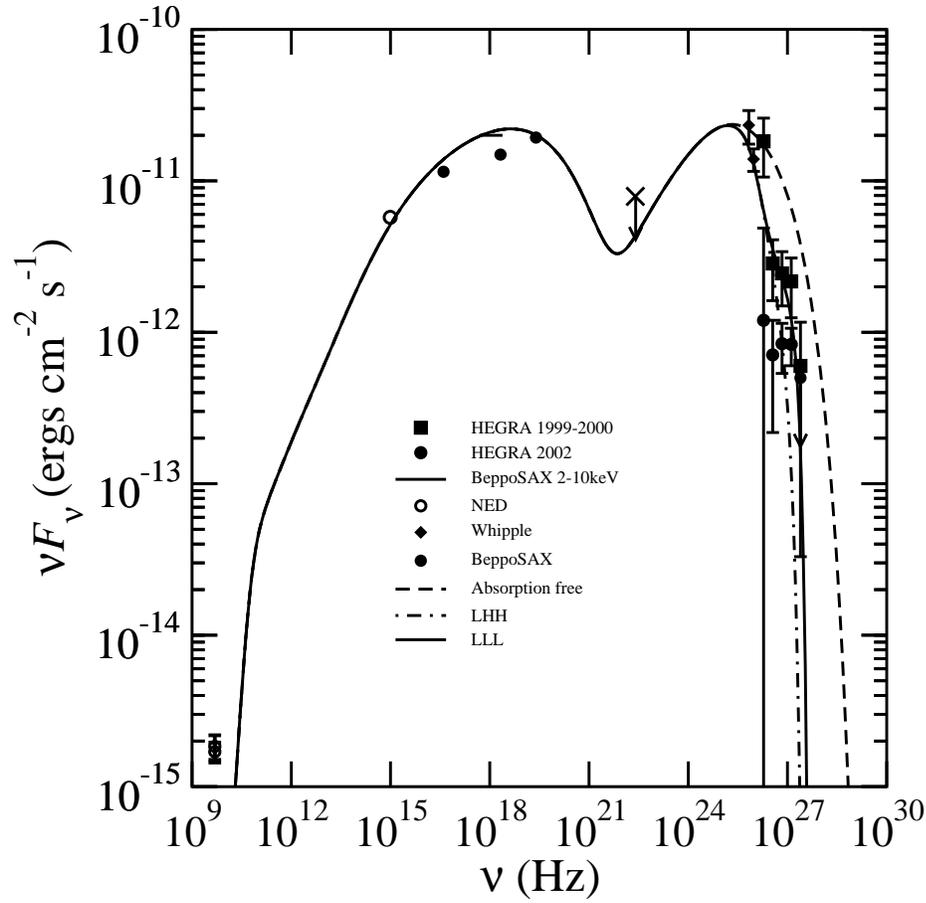}
\caption{SED of H1426+428 and model spectra with different EBL models.
Data obtained in 2002 by HEGRA are also shown for reference.
The absorption free SED model is shown by a dashed line.
The SEDs corrected by EBL models are shown by a dot-dashed line (LHH),
and a solid line (LLL).
\label{fig:h1426-sed}}
\end{figure}

Our model SEDs of H1426+428 are shown in Figure \ref{fig:h1426-all-EBL}
by solid and dashed lines.  
The parameters to specify the spectra of electrons and photons
of the emission region of H1426+428
are $\Gamma$, $R$, magnetic field $B$, 
$p$, $\gamma_\mathrm{min}$, $\gamma_\mathrm{max}$, $q_0$, 
and $t_{e, \mathrm{esc}}$.
The model spectrum of EBL that can be consistent with our SED models
are LLL, LLH, MLH, and MLL. 
With these EBL templates, it is not necessary to invoke
the external soft photons as seed photons of inverse Compton scattering, 
because of the smaller value of the optical depth for $\gamma$-rays
in 1 -- 20 TeV region.
We first adopt X = L as a template for which the model of H1426+428 
is obtained.
In Figure \ref{fig:h1426-sed}, the $\gamma$-ray data 
together with the data in other wave bands are shown,
where HEGRA data in the term 2002 are also shown for reference.
The X-ray data obtained by {\it Beppo}SAX are read from 
Figure 3 in \citet{cos03} and only three points are plotted in this figure.
It is to be noted that the data in multiwavelength
bands were not obtained simultaneously.
We find that the following values of the parameters 
for the model SED before the attenuation fit the observed data:
$B = 0.1$G, $\Gamma = 20$, $R = 10^{16}$ cm, 
$p = 1.9$, $\gamma_\mathrm{min} = 2$, $\gamma_\mathrm{max} = 10^{7}$,
the electron injection rate of $10^{-3}$ electrons cm$^{-3}$ s$^{-1}$,
and $t_{e, \mathrm{esc}} = 4 R/c$.
We call this model ``base model,'' hereafter.
(The solid line in Figure \ref{fig:h1426-all-EBL} is the SED calculated with
these parameters.)
In Figure \ref{fig:h1426-sed},
the attenuation free SED is shown by a dashed line.
The model SEDs in the observer frame are calculated by multiplying the
base model by $\exp (-\tau_{\gamma \gamma})$ with 
the EBL models, LHH and LLL.
The EBL models LHH and LHL result in almost the same SEDs,
and LLL and LLH also result in almost identical SEDs.
In these calculations, we included the pair production that
occurs in the emission region, but this effect is found to be negligible.
The peak value in the $\nu$-$\nu F_\nu$ representation of 
the SSC component is given 
at $h \nu \sim 8.5 \times 10^{10}$ eV 
and $\sim 6.5 \times 10^{10}$ eV 
for before and after the attenuation by EBL, respectively.
These values are much smaller than those obtained by \citet{cos03}.
Also these are smaller than the peak energy obtained for
Mrk 421 and Mrk 501 by \citet{dk05}.

For the base model shown in Figure \ref{fig:h1426-sed},
the energy contents of the emission region are as follows;
magnetic energy density 
$u_\mathrm{mag} = 3.98 \times 10^{-4}$ ergs cm$^{-3}$, 
nonthermal electron energy density 
$u_e = 7.51 \times 10^{-2}$ ergs cm$^{-3}$, 
and photon energy density
$u_\mathrm{ph} = 8.66 \times 10^{-2}$ ergs cm$^{-3}$.
Thus $u_e / u_\mathrm{mag} \approx 188$
and this value is larger than $u_e / u_\mathrm{mag} \sim 10$ 
for other TeV blazars \citep{kino02}.
This large value of $u_e / u_\mathrm{mag}$ is a result of
a large energy density of very high energy electrons, which is necessary
to emit a large flux of TeV $\gamma$-rays.
The high energy tail of the $\gamma$-ray spectrum
is highly affected by the Klein-Nishina effects in inverse Compton emission
and the attenuation by EBL.
If the flux of EBL is larger than LLL or LLH model,
much larger values of $u_e$ and $\Gamma$ are needed 
to account for the observed flux of TeV $\gamma$-rays.

\begin{figure}
\includegraphics[angle=0,scale=.80]{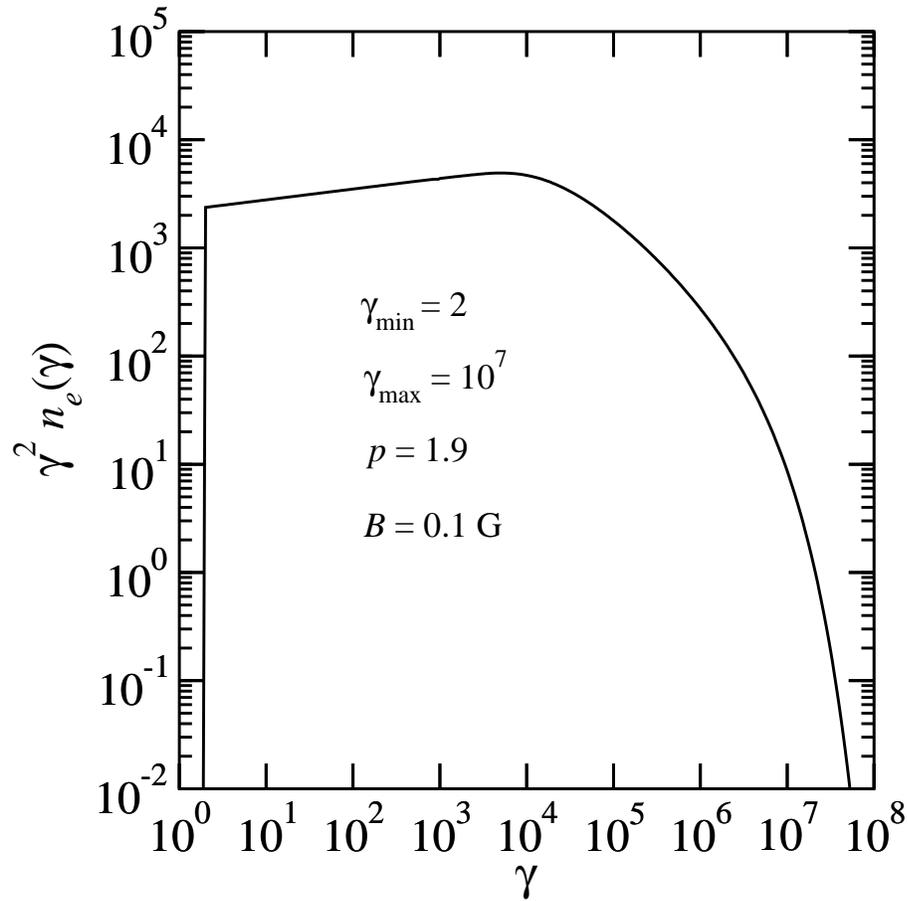}
\caption{Electron spectrum in the emission region for the base model.
The parameters for the injection spectrum are 
$\gamma_\mathrm{min} = 2$, $\gamma_\mathrm{max} = 10^7$, $p = 1.9$,
and the injection rate of $10^{-3}$ electrons cm$^{-3}$ s$^{-1}$.
\label{fig:el-spec}}
\end{figure}

The energy spectrum of nonthermal electrons is shown 
in Figure \ref{fig:el-spec} for the base model.
The injected electron spectrum has an exponential cutoff at $\gamma = 10^7$.
The steady state energy spectrum of electrons 
has a peak at $\gamma \sim 10^4$ that
is caused by the balance between radiative cooling and escape of 
electrons from the emission region.
The observed characteristic synchrotron frequency due to 
electrons with $\gamma = 10^4$ is about $8 \times 10^{14}$ Hz,
if $\Gamma = 20$ and $B = 0.1$ G.
The value of $\gamma_\mathrm{min}$ is 2 and the cooling time for electrons
with this value of $\gamma$ is so long that there is no flow in the energy
space of electrons toward lower energies below $\gamma = 2$.

\begin{figure}
\includegraphics[angle=0,scale=.80]{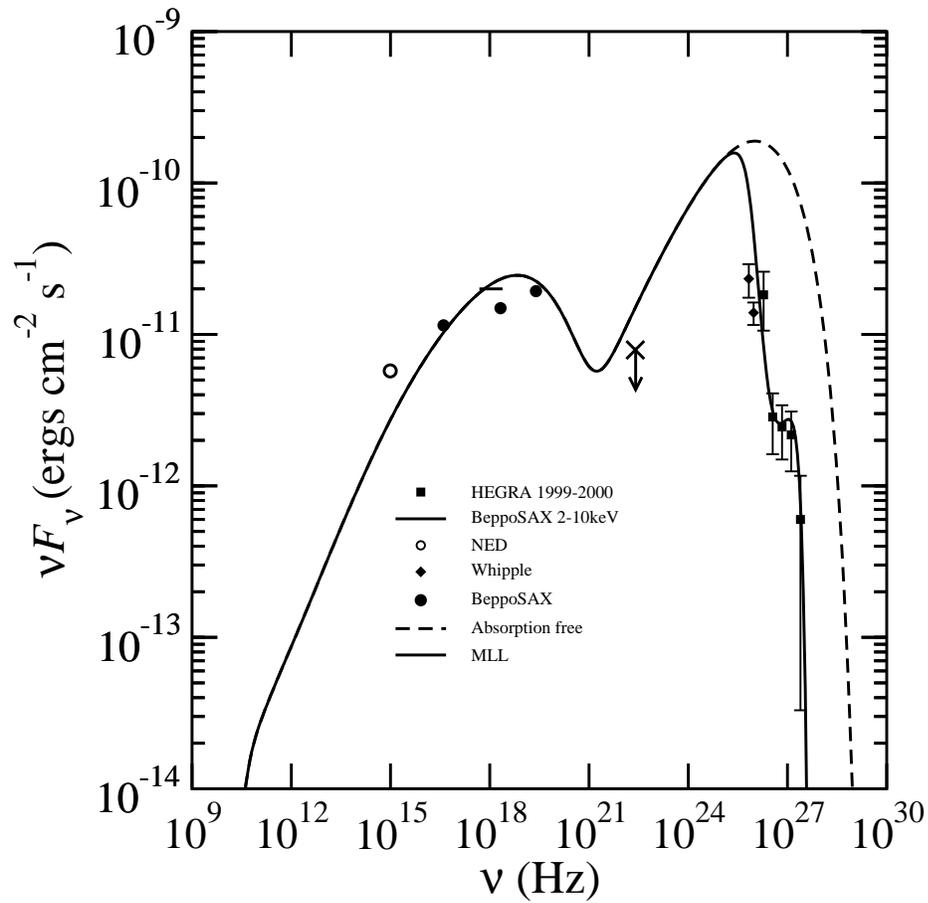}
\caption{SED of H1426+428 and a model spectrum.
The model SED corrected by an EBL model, MLL, is shown by a solid line,
and the absorption free SED model is shown by a dashed line.
\label{fig:h1426-XM-sed}}
\end{figure}

In Figure \ref{fig:h1426-XM-sed}, a model SED with X=M, i.e., MLL, is shown.
To fit the data with MLL, a very large SSC component is necessary.
For this,
$B = 0.02$ G and the injection rate of $5 \times 10^{-3}$
electrons cm$^{-3}$ s$^{-1}$ are assumed, and other parameter values
are the same as those of the base model. 
(The dashed line in Figure \ref{fig:h1426-all-EBL} is the SED calculated 
with these parameters.)
Because the SSC component highly dominates over the synchrotron component, 
an extremely large value of $u_e / u_\mathrm{mag}$ is realized, 
i.e., $\sim 2.87 \times 10^{4}$.

\begin{figure}
\includegraphics[angle=0,scale=.80]{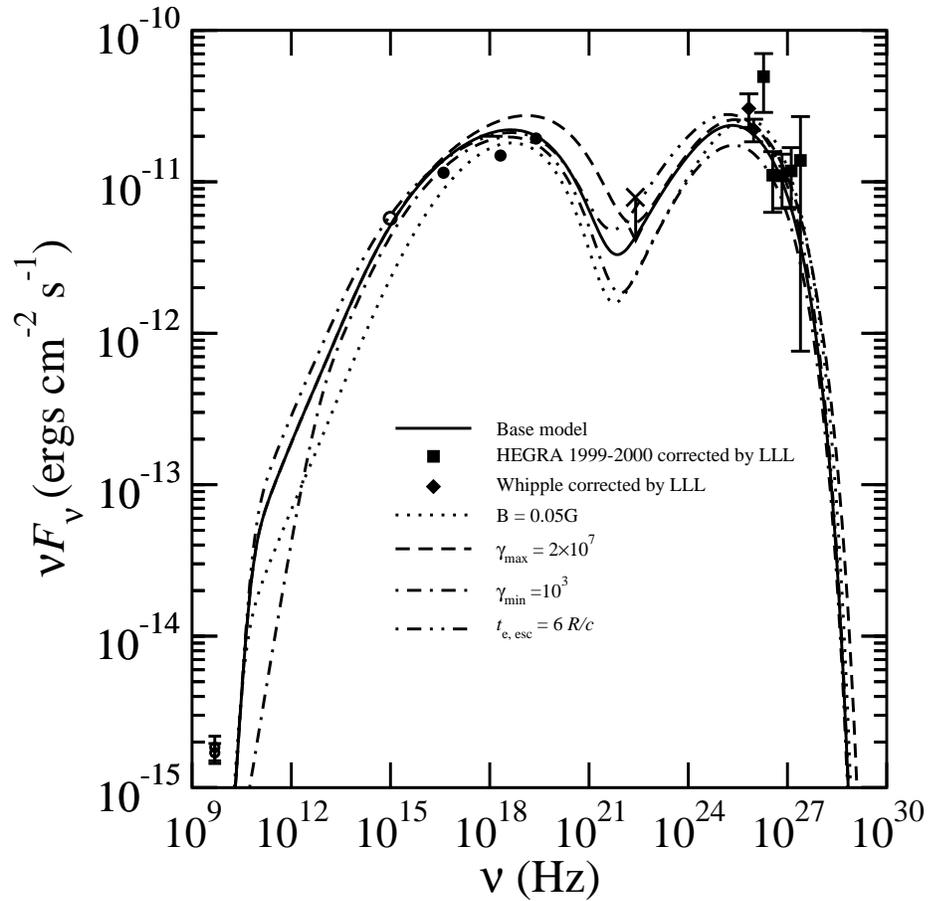}
\caption{SED models calculated with different parameter values.
Data from Whipple and HEGRA 1999-2000 are multiplied by
$\exp(\tau_{\gamma \gamma})$ with an EBL model, LLL.
Base model denotes the SED shown in Fig. \ref{fig:h1426-sed} 
by the dashed line.
\label{fig:diff-parm-sed}}
\end{figure}

We compare the models of the emission spectrum of H1426+428
with different values of parameters in Figure \ref{fig:diff-parm-sed}.
Note that in this figure observed $\gamma$-rays are corrected by multiplying
$\exp( \tau_{\gamma \gamma})$ to reproduce the SED before the attenuation,
assuming LLL model for EBL.
When the magnetic fields are weaker but the other parameter values are the same
as for the base model in Figure \ref{fig:h1426-sed},
the synchrotron cooling rate decreases and 
the number of high energy electrons increases.  
As a result the SSC component has a larger luminosity.
An example with $B = 0.05$ G is shown by a dotted line.
When the value of $\gamma_\mathrm{min}$ is larger, 
the synchrotron luminosity in radio decreases and the inverse Compton
luminosity in the MeV region also decreases.  
In Figure \ref{fig:diff-parm-sed}, $\gamma_\mathrm{min} = 10^3$
is shown by a dash-dotted line.  This SED is calculated with the value of 
$q_0$ adjusted so that the peak luminosity in X-rays is almost the same as 
that of the base model.  The difference in $t_{e, \mathrm{esc}}$ 
does not change much the spectra of X-ray and TeV $\gamma$-ray, 
although the larger value of
$t_{e, \mathrm{esc}}$ results in a larger flux in MeV $\gamma$-rays.
The important parameters to determine the TeV flux
are $\gamma_\mathrm{max}$ and $\Gamma$.  
When the value of $\gamma_\mathrm{max}$ is larger, the peak energy of
synchrotron emission becomes larger, but the peak energy of 
the SSC component does not change much because of the Klein-Nishina effects.
On the other hand, the value of $\Gamma$ does not change the spectral shape.
It appears that the base model does not yield a good fit in X-ray regime,
and the allowable parameter ranges are still large.
Since the data in X-rays and $\gamma$-rays were not obtained simultaneously
stronger constraints are difficult to set.

%%%%%%%%%%%%%%%%%%%%%%%%%%%%%%%%%%%%%%%%%%%%%%%%%%%%%%%%
%%%%%%%%%%%%%%%%%%%%%%%%%%%%%%%%%%%%%%%%%%%%%%%%%%%%%%%%

\section{Summary and Discussion}
\label{sec:sum}

We obtained the self-consistent models of the emission region of H1426+428,
taking account of the interaction of TeV photons and EBL photons via
$\gamma + \gamma \rightarrow e^+ + e^-$.
In the calculations of the emission spectrum of H1426+428,
we solved the kinetic equations of electrons and photons in a relativistically
moving blob.
To obtain the optical depth for the attenuation 
($\gamma + \gamma \rightarrow e^+ + e^-$)
we used 12 templates of EBL constructed by \citet{dk05}.

We found the parameter sets of H1426+428 for the models 
with EBL models with X = L or M and Y = L.
Here X and Y represent the energy fluxes of the stellar component
and 15 $\mu$m band, respectively.
%These SEDs of EBL are reconstructed in Figure \ref{fig:ebl-spectrum}.
Other EBL models require stronger emission
of TeV $\gamma$-rays and we did not find better models of H1426+428.
With Y = L, we found that the bulk Lorentz factor of H1426+428 is about 20 
and that the magnetic field is $\sim 0.1$ G.
These are similar values to those of other TeV blazars such as Mrk 421
\citep{kino02}.
The energy density of the nonthermal electrons dominates that of 
magnetic fields.  This is the same properties as have been found for
other TeV blazars such as Mrk 421 and Mrk 501 \citep{kino02},
but the value of $u_e/u_\mathrm{mag}$ is about 190 and this is much
larger than the values of about 10 for Mrk 421 and Mrk 501.
The injected electron number spectrum follows a power law with
a power-law index of 1.9 and has an exponential cutoff 
at $\gamma_\mathrm{max} = 10^7$.
This value of $\gamma_\mathrm{max}$ is larger than that for Mrk 421 
and Mrk 501 \citep{kino02} and
very efficient acceleration mechanisms is suggested in this source.
The electron energy spectrum in the steady state in the emission region
has a break at $\gamma \sim 10^4$.
Very efficient cooling by synchrotron
radiation and inverse Compton scattering produces this break
and a large fraction of particle energy is lost as radiation.
In many models to explain the high energy $\gamma$-rays from blazars,
the electron spectrum in the emission region is just assumed to
obey a power law or a broken-power law.  However, it is clear from
our results that solving the kinetic equation of particles is necessary
to obtain the self-consistent particle spectrum in particular when
Klein-Nishina effects are significant.

The peak value of $\nu F_\nu$ in the $\gamma$-ray region is attained
at $h \nu \sim 8.5 \times 10^{10}$ eV before the attenuation 
according to our base model.  This value is significantly smaller
than the peak energy estimated by \citet{cos03} for H1426+428 
(larger than 10 TeV) and
the values estimated for Mrk 421 and Mrk 501 ($\sim 1$-3 TeV) by \citet{dk05}.
\citet{cos03} concluded that inverse Compton scattering of external
soft photons produces TeV $\gamma$-rays, because SSC models suffer
from the Klein-Nishina effects.  
However, our results show that there is no need for 
the external soft photon sources.  
This discrepancy originates from the use of different models of EBL.  
\citet{cos03} used the models by \citet{aha02} and \citet{pri01} 
and these models yield higher intensity of EBL than LLL and LLH.
Then the similar arguments might be raised,
if EBL models such as HHH, HHL, MHH, and MHL are applied.
That is, for these EBL models, 
TeV emission appears to have a peak above 50 TeV in SED before the attenuation
and inverse Compton scattering of external soft photons might 
be necessary to account for the $\gamma$-rays spectrum.
If the X-ray emission at $\sim 10^{19}$ Hz (a peak frequency in SED)
is by synchrotron emission,
radiating electrons have Lorentz factor of $\sim 10^6$, 
where $B = 0.1$ G and $\Gamma = 20$ are assumed.
To produce $\gamma$-rays with energy of 50 TeV via inverse Compton scattering
by the same electrons in the Thomson regime, 
the energy of soft photons in the rest frame of external
soft photon sources is in the IR region with the frequency of
about $10^{13}$ Hz and the broad line regions can be a photon source.
Because of the large cooling rate of Compton scattering of 
external soft photons,
the values of $B$, $\gamma_\mathrm{max}$, and $q_0$ will be modified.
It is, however, beyond the scope of this paper to find another model
with Compton scattering of external soft photons.

\citet{dk05} analyzed the data of Mrk 421, Mrk 501, and H1426+428,
and they rejected some EBL spectra which cause ``unphysical intrinsic spectra
characterized by an exponential rise at high TeV energies.''
The EBL models, LLL, LLH, MLL and MLH, 
which are found to be good candidates for
the EBL spectrum in our model, are also allowed models in \citet{dk05}.
Since they only tested the behavior of high energy $\gamma$-rays,
their constraints may not be restrictive.
In contrast, we searched for a set of parameter values that
realizes multiwavelength properties of H1426+428, based on 
the homogeneous SSC model.

In this paper we assumed that the phase space density of EBL is
constant [equation (\ref{eq:ebl-local})]. 
It is possible that the EBL spectrum may not change with redshift so that
$n_\mathrm{EBL}(\nu, z) d \nu = n_0(\nu_0) d \nu_0$,
because stellar population and dust emission might be almost homogeneous
for small values of $z$.  We tested this model of EBL and found 
the value of $\tau_{\gamma \gamma}$  of LLL at $E_0 = 1$ TeV decreases
about 18 \%.
Then above mentioned results with the constant phase space density 
of EBL are not changed much.

Because the multiwavelength data of H1426+428 have not been obtained 
simultaneously,
the constraints on the physical condition of H1426+428
and the EBL spectrum are not stringent yet.
Also, because there are several parameters to determine the emission spectrum
from the jet, the allowed ranges of the parameter values are
not strictly determined.  According to our numerical calculations,
the magnetic fields, for example, might be in
the range between 0.05 and 0.2 G, and the bulk Lorentz factor might be
in the range between 15 and 25.  
However, the properties of H1426+428 found in this work will not be changed 
dramatically as long as the SSC model is applied
and the EBL flux at around 10 $\mu$m should be as low as the model values
such as LLL, LLH, MLL, and MLH.

Recent observations of PKS 2155--304 by H.E.S.S. (High Energy Stereoscopic 
System) showed that it emits
$\gamma$-rays above 160 GeV and possibly above a few TeV \citep{aha05}.
Because this object is located at $z = 0.116$, 
the absorption of TeV $\gamma$-rays
by EBL is probable and further constraints on the EBL spectrum
will be obtained.
More recently, \citet{aha05astroph} reported the detection of TeV
$\gamma$-rays by H.E.S.S. from further distant blazars, i.e.,
H2356--309 ($z = 0.165$) and 1ES 1101--232 ($z = 0.186$).
Assuming that the source spectra of $\gamma$-rays obey a power law,
$dN/dE \propto E^{-\alpha}$, where $dN/dE$ is the differential number
spectrum of $\gamma$-rays and $\alpha$ is the power-law index,
they set upper limits on the EBL spectrum
in the wavelength band $\sim 0.1$ -- 10 $\mu$m.
The EBL models of LLL and LLH that are favored in this work 
are consistent with the EBL models 
of P0.45 and P0.55 of \citet{aha05astroph} in $\lambda = 1$ -- 10 $\mu$m.

\citet{dka05} recently examined the extragalactic nature of the excess
near-infrared background light (NIRBL) \citep[e.g.,][]{mat05}, 
using the analysis of the $\gamma$-ray
spectra of blazars such as H1426+428 and PKS 2155--304.
This excess NIRBL is still controversial and they argued that the excess
NIRBL may not be related with the spectral imprint of the first
generation of stellar objects on the EBL.

In this paper we assumed the steady state of the emission region,
but the time variation of X-rays and TeV $\gamma$-rays of blazars
is common.
The time evolution of the emission spectrum
might be used to constrain the emission mechanisms of blazars including
external soft photons and hadronic models.  Then the EBL spectrum
might be constrained further.
Since our numerical scheme is capable of
time-dependent calculations, we may apply our model to 
time variable blazar emission in future work.

%%%%%%%%%%%%%%%%%%%%%%%%%%%%%%%%%%%%%%%%%%%%%%%%%%%%%%%%%%%%%%
\acknowledgments
This work has been partially supported by Scientific Research Grants 
(M.K.: 15037210; F.T.: 14079205 and 16540215) from 
the Ministry of Education, Culture, Sports, Science and Technology of Japan.

%%%%%%%%%%%%%%%%%%%%%%%%%%%%%%%%%%%%%%%%%%%%%%%%%%%%%%%%%%%%%%

\clearpage

\begin{thebibliography}{}
\bibitem[Aharonian et al. (2002)]{aha02}
Aharonian, F. A., et al. 2002, \aap, 384, L23

\bibitem[Aharonian et al. (2003)]{aha03}
Aharonian, F. A., et al. 2003, \aap, 403, 523

\bibitem[Aharonian et al. (2005a)]{aha05}
Aharonian, F. A., et al. 2005a, \aap, 430, 865

\bibitem[Aharonian et al. (2005b)]{aha05b}
Aharonian, F. A., et al. 2005b, \aap, 441, 465

\bibitem[Aharonian et al. (2005c)]{aha05astroph}
Aharonian, F. A., et al. 2005c, astro-ph/0508073

\bibitem[Bennett et al. (2003)]{ben03}
Bennett, C. L., et al. 2003, \apjs, 148, 1

\bibitem[Costamante et al. (2001)]{cos01}
Costamante, L., et al. 2001, \aap, 371, 512

\bibitem[Costamante et al. (2003)]{cos03}
Costamante, L., Aharonian, F., Ghisellini, G., \& Horns, D. 2003, 
New Astronomy, 47, 677

\bibitem[Coppi \& Blandford (1990)]{cb90} 
Coppi, P. S., \& Blandford, R. D. 1990, \mnras, 245, 453 

\bibitem[De Jager \& Stecker (2002)]{djs02}
De Jager, O. C., \& Stecker, F. W. 2002, \apj, 566, 738

\bibitem[Djannati-Atai et al. (2002)]{dja02}
Djannati-Atai, A., et al. 2002, \aap, 391, L25

\bibitem[Dwek \& Krennrich (2005)]{dk05}
Dwek, E., \& Krennrich, F. 2005, \apj, 618, 657

\bibitem[Dwek \& Slavin (1994)]{ds94}
Dwek, E., \& Slavin, J. 1993, \apj, 436, 696

\bibitem[Dwek et al. (2005)]{dka05}
Dwek, E., Krennrich, F., \& Arendt, R. G., astro-ph/0508133, to appear in \apj

\bibitem[Falcone et al. (2004)]{fal04}
Falcone, A. D., Cui, W., \& Finley, J. P. 2004, \apj, 601, 165

\bibitem[Gould \& Schr\`{e}der (1966)]{gs66}
Gould, R. J., \& Schr\`{e}der, G. 1966, \prl, 16, 252

\bibitem[Hartman et al. (1992)]{har92}
Hartman, R., et al. 1992, \apjl, 385, L1

\bibitem[Hauser \& Dwek (2001)]{hd01}
Hauser, M. G., \& Dwek, E. 2001, \araa, 39, 249

\bibitem[Horan et al. (2002)]{hor02}
Horan, D., et al. 2002, \apj, 571, 753

\bibitem[Inoue \& Takahara (1996)]{it96} 
Inoue, S., \& Takahara, F. 1996, \apj, 463, 555 

\bibitem[Jelly (1966)]{jel66}
Jelly, J. V. 1966, \prl, 16, 479

\bibitem[Kashlinsky (2005)]{kas05}
Kashlinsky, A. 2005, \physrep, 409, 361

\bibitem[Kino et al. (2002)]{kino02}
Kino, M., Takahara, F., \& Kusunose, M. 2002, \apj, 564, 97

\bibitem[Kneiske et al. (2004)]{kne04}
Kneiske, T. M., Bretz, T., Mannheim, K., \& Hartmann, D. 2004, \aap, 413, 807

\bibitem[Konopelko et al. (2003)]{kon03}
Konopelko, A., et al. 2003, \apj, 597, 851

\bibitem[Kusunose et al. (2003)]{kus03}
Kusunose, M., Takahara, F., \& Kato, T. 2003, \apjl, 592, L5

\bibitem[Li \& Kusunose (2000)]{lk00}
Li, H., \& Kusunose, M. 2000, \apj, 536, 729

\bibitem[Malkan \& Stecker (2001)]{ms01}
Malkan, M. A., \& Stecker, F. W. 2001, \apj, 555, 641

\bibitem[Mannheim \& Biermann (1992)]{mb92}
Mannheim, K., \& Biermann, P. 1992, \aap, 253, L21

\bibitem[Mannheim (1993)]{man93}
Mannheim, K. 1993, \aap, 269, 67

\bibitem[Maraschi et al. (1992)]{mgc92}
Maraschi, L., Ghisellini, G., \& Celotti, A. 1992, \apjl, 397, L5

\bibitem[Marscher \& Travis (1996)]{mt96}
Marscher, A. P., \& Travis, J. P. 1996, \aap, Suppl., 120, 537

\bibitem[Matsumoto et al. (2005)]{mat05}
Matsumoto, T., et al. 2005, \apj, 626, 31

\bibitem[M\"{u}cke et al. (2003)]{muc03}
M\"{u}cke, A., et al. 2003, Astropart. Phys., 18, 593

\bibitem[Nikishov (1962)]{nik62} 
Nikishov, A. I. 1962, Sov. Phys. JETP, 14, 393

\bibitem[Petry et al. (2002)]{pet02}
Petry, D., et al. 2002, \apj, 580, 104

\bibitem[Primack et al. (2001)]{pri01}
Primack, J. R., et al. 2001, AIP Conf. Proc., 558, 463 (New York: AIP)

\bibitem[Stecker et al. (1992)]{sjs92} 
Stecker, F. W., De Jager, O. C., \& Salamon, M. H. 1992, \apjl, 390, L49

\bibitem[Stecker \& Salamon (1996)]{ss96} 
Stecker, F. W., \& Salamon, M. H. 1996, \apj, 464, 600
\end{thebibliography}
\end{document}